\documentclass[12pt]{article}
\usepackage{times,mathptm}
\usepackage[dvips]{graphicx}
\usepackage{epsfig}
\usepackage{subfigure}
\usepackage{natbib}
\usepackage{afterpage}
\renewcommand{\author}[1]{\noindent  #1\\}
\newcommand{\address}[1]{ {\small \noindent #1\\}}
\renewcommand{\title}[1]{{{\Large \noindent \bf#1\\}}}
\renewcommand{\baselinestretch}{2.0}
\def\spose#1{\hbox to 0pt{#1\hss}}
\def\lta{\mathrel{\spose{\lower 3pt\hbox{$\mathchar"218$}}
     \raise 2.0pt\hbox{$\mathchar"13C$}}}
\def\gta{\mathrel{\spose{\lower 3pt\hbox{$\mathchar"218$}}
     \raise 2.0pt\hbox{$\mathchar"13E$}}}

\def\aap{{\em Astron. Astrophys. \ }}
\def\aaps{{\em Astron. Astrophys. Suppl. \ }}
\def\apj{{\em Astrophys. J. \ }}
\def\apjl{{\em Astrophys. J. Lett. \ }}
\def\grl{{\em Geophys. Res. Lett. \ }}
\def\ija{{\em Int. J. Astrobiology \ }}
\def\jgr{{\em J. Geophys. Res. \ }}
\def\mnras{{\em Mon. Not. Roy. Astron. Soc. \ }}

\begin{document}
\begin{center}
\title{\bf Hypothesis Paper}
\vspace{0.75cm}
\title{Habitability of Super-Earth Planets around Other Suns: \\
Models including Red Giant Branch Evolution}

\vspace{1.25cm}
\author{W. von Bloh$^1$, M. Cuntz$^2$, K.-P. Schr\"oder$^3$, C. Bounama$^1$, S. Franck$^1$}
\vspace{1.0cm}
\address{$^1$Potsdam Institute for Climate Impact Research,
14412 Potsdam, Germany \\
email: bloh@pik-potsdam.de, bounama@pik-potsdam.de, franck@pik-potsdam.de}
\vspace{0.5cm}
\address{$^2$Department of Physics, University of Texas at
Arlington, Arlington, TX 76019, USA \\
email: cuntz@uta.edu}
\vspace{0.5cm}
\address{$^3$Department of Astronomy, University of Guanajuato,
36000 Guanajuato, GTO, Mexico \\
email: kps@astro.ugto.mx}
\vspace{2.0cm}
Submitted to \enskip {\em{Astrobiology}}\\
\vspace{1.0cm}
Submitted \enskip \today\\
\vspace{1.0cm}
\end{center}
\begin{flushleft}
{\bf{Running Title:}} \enskip Habitability of Super-Earths including RGB evolution\\
\begin{flushleft} {\bf Key words:} 
Extrasolar terrestrial planets --- Habitable zone --- Planetary atmospheres ---
Modelling studies.\end{flushleft}

\pagebreak

{\bf{Corresponding Author:}} \\
\vspace{1.0cm}
Dr. Werner von Bloh\\
Potsdam Institute for Climate Impact Research \\
Box  60 12 03\\
14412 Potsdam \\
Germany \\
\vspace{1.0cm}
Phone: +49-331-288-2603 \\
Fax: +49-331-288-2600 \\
E-mail: bloh@pik-potsdam.de
\end{flushleft}
\pagebreak

\noindent
{\bf ABSTRACT}

\bigskip
\noindent
The unexpected diversity of exoplanets includes a growing number of super-Earth planets,
i.e., exoplanets with masses of up to several Earth masses and a similar chemical and
mineralogical composition as Earth.  We present a thermal evolution model for a
10 Earth mass planet orbiting a star like the Sun.  Our model is based on the
integrated system approach, which describes the photosynthetic biomass production
taking into account a variety of climatological, biogeochemical, and geodynamical
processes.  This allows us to identify a so-called photosynthesis-sustaining habitable
zone (pHZ) determined by the limits of biological productivity on the planetary surface.
Our model considers the solar evolution during the main-sequence stage and along the
Red Giant Branch as described by the most recent solar model.  We obtain a large set
of solutions consistent with the principal possibility of life.  The highest likelihood
of habitability is found for ``water worlds''.  Only mass-rich water worlds are able to
realize pHZ-type habitability beyond the stellar main-sequence on the Red Giant Branch.

\pagebreak

\section{Introduction}

\bigskip
\noindent
A pivotal part in the ongoing search for extra-solar planets is the quest to
identify planetary habitability, i.e., the principal possibility of life.
In a previous paper, \cite{kast93} presented a one-dimensional climate model
to define a zone of habitability (HZ) around the Sun and other main-sequence
stars that assumed as basic premise an Earth-like model planet with a
CO$_2$/H$_2$O/N$_2$ atmosphere and that habitability requires the presence
of liquid water on the planetary surface.    

In the meantime, other definitions of habitable zones have been proposed
such as the Galactic HZ, the UV-HZ, and the photosynthesis-sustaining HZ
(pHZ).  The Galactic HZ \citep{line04} caters to the requirement that a
sufficient amount of heavy elements (notably those contained in carbon and
silicate compounds) must be present for the build-up of planets and life, a
condition easily met in the solar neighborhood.  The UV-HZ \citep{bucc06,cunt08}
is based on the premise that no lethal amounts of stellar UV flux is produced
(regarding life forms assuming carbon-based biochemistry), a condition that
tends to favor the environment of old main-sequence stars and giants
\citep{guin02} as well as planets with appreciable atmospheres, notable
significant ozone layers \citep{segu03}.

Another definition of habitability first introduced by \cite{fran00a,fran00b}
is associated with the photosynthetic activity of the planet, which
critically depends on the planetary atmospheric CO$_2$ concentration.
This type of habitability is thus strongly influenced by the planetary
geodynamics, encompassing climatological, biogeochemical, and geodynamical
processes (``Integrated System Approach'').  This concept has previously
been used in studies of fictitious planets around 47~UMa \citep{cunt03,fran03}
and 55~Cnc \citep{bloh03}, as well as detailed studies of observed super-Earth
planets in the Gliese 581 system \citep{bloh07b}.  The latter investigation
showed that Gliese 581c is clearly outside the habitable zone, since it is
too close to the star, whereas Gliese 581d located near the outer edge of the
habitable zone is probably habitable, at least for certain types
of primitive life forms \citep[see also][]{sels07}.  Moreover,
\cite{bloh07a} have used this type of
model to compile a detailed ranking of known star-planet systems regarding
the principal possibility of life, which by the way led to the conclusion
that the Solar System is not the top-tier system (``Principle of Mediocrity'').

In case of Earth-mass planets (1~$M_\oplus$), a detailed investigation of
geodynamic habitability was presented by \cite{fran00b} with respect
to the Sun as well as stars of somewhat lower and higher mass as central
stars.  \citeauthor{fran00b} found that Earth is rendered uninhabitable
after 6.5~Gyr as a result of plate tectonics, notably the growth of the
continental area (enhanced loss of atmospheric CO$_2$ by the increased
weathering surface) and the dwindling spreading rate (diminishing CO$_2$
output from the solid Earth). 

This implies that there is no merit in investigating the future habitability
of Earth during the post--main-sequence evolution of the Sun, as in the
framework of pHZ models, the lifetime of habitability is limited by
terrestrial geodynamic processes.  However, this situation is expected
to be significantly different for super-Earth planets due to inherent
differences compared to Earth-mass planets \citep[e.g.,][]{vale07a}.
A further motivation for this type of work stems from the ongoing discovery
of super-Earths in the solar neighborhood with the Gliese 876 \citep{rive05}
and Gliese 581 \citep{udry07} systems as prime examples. 

In the following, we discuss the definition of the photosynthesis-sustaining
habitable zone, including the relevant geodynamic assumptions.  Next, we
describe the most recent model of solar evolution that is used as basis
for our study.  Thereafter, we present our results including comparisons to
previous work.  Finally, we present our summary and conclusions.

\section{A New Model of Solar Evolution}


\bigskip
\noindent
A key element of the present study is to consider a star akin to the
Sun as the central object of the star-planet system.  \cite{schr08} recently
obtained a new model of solar evolution that will be adopted in the
following.  This model is based on a well-tested stellar evolution code
that allows us to follow the change of solar properties at the main-sequence
(MS), along the Red Giant Branch (RGB) and beyond.  It is the Eggleton
evolution code in the version described by \cite{pols95,pols98}, which
has updated opacities and an improved equation of state.

Among other desirable characteristics, the code uses a self-adapting mesh
and a proper treatment of ``overshooting" that has been tested and calibrated
with giant and supergiant stars in eclipsing binary systems.  The code also
considers a detailed description of the mass loss following \cite{schr05}
that has been tested based on a set of well-observed stars \citep{schr07}. 
Thus it permits an accurate description of the time-dependent solar luminosity
along the RGB (see Fig.~1).  A further consequence of the steadily increasing
mass loss is the increase of the orbital distances $R$ of any putative planets,
given as $R \propto M_\odot^{-1}$ with $M_\odot$ as solar mass owing
to the conservation of the orbital angular momentum of the planet.

The solar evolution model by \cite{schr08} suggests an age of the Sun as
4.58 ($\pm 0.05$) Gyr, and the RGB-tip is reached after 12.167 Gyr, which is
also the point in time, where our computations are suspended.  This model
also confirms some well-established facts: (1) The MS-Sun has already undergone
significant changes, i.e., the present solar luminosity $L_\odot$ exceeds the
zero-age value by 0.30 $L_\odot$, and the zero-age solar radius has been
11\% smaller than the present value.  (2) There was an increase of effective
temperature from 5596~K to 5774 ($\pm 5$) K. (3) The present Sun is increasing
its average luminosity at a rate of 1\% in every 110 million years, or 10\%
over the next billion years.  All these findings are consistent with
established solar models like the one by \cite{goug81} and subsequent work.

During the solar MS, the consequences of evolution for Earth-type planets
(as well as other types of planets) are extremely slow, compared to natural
or human-driven climate changes on Earth.  Nonetheless, solar-type
evolution will force global warming upon any planet, which has been the
subject of detailed previous investigations both concerning the climatic
HZ \citep[e.g.,][]{unde03,jone05} and the photosynthesis-sustaining habitable
zone of the Sun \citep[e.g.,][]{fran00b}.  According to the evolution model
by \cite{schr08}, the tip-RGB evolution will be reached with a luminosity of
2730 $L_\odot$, an effective temperature of 2602~K, and a radius of 256~$R_\odot$.
At that time, the Sun will have lost 0.332~$M_\odot$ of its initial mass.
There is an ongoing debate at what point in time a planet
originally located at 1~AU, equating 215~$R_\odot$, will be engulfed as a
consequence.  Contrary to the previous model by \cite{sack93} that is
based on a less accurate description of the solar mass loss, \cite{schr08}
concluded that such an engulfment will happen during the late phase of the
solar RGB evolution.  In fact, the minimal orbital radius for a planet able
to survive is found to be about 1.15~AU.

The evolution of the central star, as well as its effects on planetary
orbits, has significant consequences for planetary habitability.
This property has previously been investigated for different types
of climatic habitable zones by \cite{kast93}, \cite{unde03}, \cite{jone05},
and others.  Furthermore, a previous assessment of the spatial and temporal
evolution of climatic HZs for different types of stars beyond the
main-sequence has been given by \cite{lope05}.  They showed that
for a 1~$M_\odot$ star at the first stages of its post--main-sequence
evolution, the temporal transit of the HZ is estimated to be several times
$10^9$~yr at 2~AU and about $10^8$~yr at 9~AU.  \cite{lope05} concluded
that under these circumstances life could develop at distances in the range
of 2 to 9~AU in the environment of subgiant or giant stars.  This view is
consistent with our current understanding that terrestrial life existed
at least as early as $7 \times 10^8$~yr after the Earth formed, which tends
to imply that life may be able to form over time intervals from $5 \times 10^8$
to $10^9$~yr. The short-time window ($\approx 10^8$~yr) for the origin of life
is bounded by the last ocean-vaporizing impact and the
earliest evidence for life on Earth ($\approx 3.8 - 3.9\times 10^9$~yr ago).
This window might be extended if the origin of life occurred close to
$3.5\times 10^9$~yr ago \citep{chyb05}.

The main goal of this study is to investigate habitability in the framework
of the photosynthesis-sustaining HZ for stars like the Sun with special
consideration of the post--main-sequence evolution.
Our study will be focused on super-Earth planets, and we will consider a
significant set of geodynamic processes.  Our findings will also be compared
with the previous work by \cite{lope05}.

\clearpage

\section{Habitability of Super-Earths}

\subsection{Definition of the photosynthesis-sustaining habitable zone}


To assess the habitability of terrestrial planets, including super-Earth planets,
an Earth-system model is applied to calculate the evolution of the temperature
and atmospheric CO$_2$ concentration.  On Earth, the carbonate-silicate cycle
is the crucial element for a long-term homeostasis under increasing solar
luminosity.  On geological time-scales, the deeper parts of the Earth are
considerable sinks and sources of carbon.  The role of weathering for the Earth's
climate was first described by \cite{walk81}.  They found that an increase in
luminosity leads to a higher mean global temperature causing an increase in
weathering.  As a consequence, more CO$_2$ is extracted from the atmosphere, thus
weakening the greenhouse effect.  Overall the temperature is lowered and homeostasis
is achieved.

On geological time scales, however, the deeper parts of the Earth are considerable
sinks and sources for carbon.  As a result, the tectonic activity and the continental
area change considerably.  Therefore, \citet{taji92} have favored the so-called
``global carbon cycle''.  In addition to the usual carbonate-silicate geochemical
cycle, it also contains the subduction of large amounts of carbon into the mantle
with descending slabs and the degassing of carbon from the mantle at mid-ocean ridges.
In particular, the potential of weathering to stabilize the surface temperature of
a terrestrial planet by a negative feedback mechanism is also strongly modulated
by the biosphere.

Our numerical model couples the solar luminosity $L$, the silicate-rock
weathering rate $F_{\mathrm{wr}}$, and the global energy balance to obtain
estimates of the partial pressure of atmospheric carbon dioxide
$P_{\mathrm{CO}_2}$, the mean global surface
temperature $T_{\mathrm{surf}}$, and the biological productivity $\Pi$ as
a function of time $t$ (Fig.~2).  The main point is the
persistent balance between the CO$_2$ (weathering) sink in the atmosphere-ocean system and
the metamorphic (plate-tectonic) sources.  This is expressed through the
dimensionless quantities  
\begin{equation}
f_{\mathrm{wr}}(t) \cdot f_A(t) \ = \ f_{\mathrm{sr}}(t),
\label{gfr}
\end{equation}
where $f_{\mathrm{wr}}(t) \equiv F_{\mathrm{wr}}(t)/F_{\mathrm{wr},0}$ is the 
weathering rate, $f_A(t) \equiv A_c(t)/A_{c,0}$ is the continental area, and
$f_{\mathrm{sr}}(t) \equiv S(t)/S_0$ is the areal spreading rate, which are all
normalized by their present values of Earth.  
Eq.~(\ref{gfr}) can be rearranged by introducing the geophysical forcing ratio
GFR \citep{volk87} as
\begin{equation}
f_{\mathrm{wr}}(T_{\mathrm{surf}},P_{\mathrm{CO}_2}) \ = \
\frac{f_{\mathrm{sr}}}{f_A} \ =: \ \mathrm{GFR}(t) .
\label{gfr2}
\end{equation}
Here we assume that the weathering rate depends only on the global surface
temperature and the atmospheric CO$_2$ concentration.  For the investigation
of a super-Earth under external forcing, we adopt a model planet with a prescribed
continental area.  The fraction of continental area relative to the total planetary
surface $c$ is varied between $0.1$ and $0.9$.

The connection between the stellar parameters and the planetary climate can be
formulated by using a radiation balance equation 
\begin{equation}
\frac{L}{4\pi R^2} [1- a (T_{\mathrm{surf}}, P_{\mathrm{CO}_2})]
 \ = \ 4I_R (T_{\mathrm{surf}}, P_{\mathrm{CO}_2}),
\label{L}
\end{equation}
where $L$ denotes the stellar luminosity, $R$ the planetary distance,
$a$ the planetary albedo, and $I_R$ the outgoing infrared flux of the planet.
Following \cite{will98} $I_R$ has been approximated by a third order polynomial and
$a$ by a second order polynomial.  These approximations have been derived from
$24,000$ runs of a radiation-convection model by \cite{kast86} and \cite{kast88}.
They are valid in a range of $10^{-9}~\mathrm{bar}<P_{\mathrm{CO}_2}<10~\mathrm{bar}$.
The Eqs.~(\ref{gfr2}) and (\ref{L}) constitute a set of two coupled equations with
two unknowns, $T_{\mathrm{surf}}$ and $P_{\mathrm{CO}_2}$, if the parameterization
of the weathering rate, the luminosity, the distance to the central star and the
geophysical forcing ratio are specified.  Therefore, a numerical solution can be
attained in a straightforward manner.

The photosynthesis-sustaining HZ is defined as the spatial domain of all distances $R$ from
the central star, e.g., the Sun, where the biological productivity is greater than zero, i.e.,
\begin{equation}
{\mathrm{pHZ}} \ := \ \{ R \mid \Pi (P_{\mathrm{CO}_2}(R,t), T_{\mathrm{surf}}(R,t))>0 \}.
\label{hz}
\end{equation}
In our model, biological productivity is considered to be solely a function of
the surface temperature and the CO$_2$ partial pressure in the atmosphere.
Our parameterization yields maximum productivity at $T_{\mathrm{surf}} = 50^{\circ}$C
and zero productivity for $T_{\mathrm{surf}} \leq 0^{\circ}$C or $T_{\mathrm{surf}}
\geq 100^{\circ}$C or $P_{\mathrm{CO}_2}\leq 10^{-5}$ bar \citep{fran00a}.
A photosynthesis-based biosphere of a super-Earth may, however, use methane to
produce CO$_2$, because hydrogen is less likely to escape to space.
The inner and outer boundaries of the pHZ do not depend on
the detailed parameterization of the biological productivity within the temperature
and pressure tolerance window.  Hyperthermophilic life forms can tolerate
temperatures somewhat above $100^{\circ}$C.  However, these chemoautotrophic organisms
are outside the scope of this study.

\subsection{Silicate Rock Weathering}

Weathering plays an important role in Earth's climate because it provides the main sink for
atmospheric carbon dioxide. The overall chemical reactions for the weathering process are
\begin{eqnarray*}
\mbox{CO$_2$}+ \mbox{CaSiO$_3$} &\rightarrow & \mbox{CaCO$_3$} + \mbox{SiO$_2$}, \\
\mbox{CO$_2$}+ \mbox{MgSiO$_3$} & \rightarrow & \mbox{MgCO$_3$} + \mbox{SiO$_2$}.
\end{eqnarray*}
The total process of weathering embraces (1) the reaction of silicate minerals with
carbon dioxide, (2) the transport of weathering products, and (3) the deposition of
carbonate minerals in the oceanic crust.  The available thickness of crust where
CaCO$_3$ is stable in the presence of silicate scales inversely with the thermal gradient
and hence inversely with surface gravity.  Therefore, there may be a problem for
storing carbonates in the crust of super-Earth planets.  Additionally, there is
an exchange with the mantle via alteration of the oceanic crust.

When combining all these effects, the normalized global mean weathering rate $f_{\mathrm{wr}}$ can be
calculated as
\begin{equation}
f_{\mathrm{wr}} \ = \ {\left( \frac{a_{\mathrm{H}^+}}{a_{\mathrm{H}^+,0}}\right)}^{0.5}
     \exp \left( \frac{T_{\mathrm{surf}}-T_{\mathrm{surf},0}}{13.7~\mathrm{K}} \right) \label{hz:eq1}
\end{equation}
following \cite{walk81}.
Here the first factor reflects the role of the CO$_2$ concentration in the soil, $P_{\mathrm{soil}}$,
with $a_{\mathrm{H}^+}$ as the activity of $\mathrm{H}^+$ in fresh soil-water that depends on
$P_{\mathrm{soil}}$ and the global mean surface temperature $T_{\mathrm{surf}}$. The quantities
$a_{\mathrm{H}^+,0}$ and $T_{\mathrm{surf},0}$ are the present-day values for the $\mathrm{H}^+$
activity and the surface temperature, respectively.
The activity $a_{\mathrm{H}^+}$ is itself a function of the temperature and the CO$_2$ concentration of the soil.
The concentration of CO$_2$ in the soil water [CO$_2$(aq)] can be obtained from the partial pressure of CO$_2$ in the soil according
to
\begin{equation}
[\mathrm{CO}_2(aq)] \ = \ K_{\rm H} P_{\mathrm{soil}}
\end{equation}
where $K_{\rm H}$ is Henry's law constant. We assume that
[CO$_2$(aq)] = [H$_2$CO$_3^\ast$]. H$_2$CO$_3^\ast$ dissociates in two steps
which are
\begin{eqnarray*}
\mathrm{H}_2\mathrm{CO}_3^\ast & \rightarrow  &\mathrm{H}^++\mathrm{HCO}_3^- \\
\mathrm{HCO}_3^- & \rightarrow & \mathrm{H}^++\mathrm{CO}_3^{2-} 
\end{eqnarray*}
The corresponding concentrations can be calculated from the law of masses as 
\begin{eqnarray}
[\mathrm{H}\mathrm{CO}_3^-] & = & \frac{K_1}{[\mathrm{H}^+]}K_{\rm H} P_{\mathrm{soil}},\label{bal1}\\ 
 \left [ \mathrm{CO}_3^{2-}\right ] & = & \frac{K_1K_2}{[\mathrm{H}^+]^2}K_{\rm H} P_{\mathrm{soil}}, \label{bal2}
\end{eqnarray}
where $K_1$ and $K_2$ are (temperature dependent) equilibrium constants. An additional
constraint for the concentrations is given by the charge balance
\begin{equation}
[\mathrm{H}^+] \ = \ [\mathrm{HCO}_3^-]+2[\mathrm{CO}_3^{2-}]+[\mathrm{OH}^-].\label{bal3}
\end{equation}
Here [H$^+$] and therefore $a_{\mathrm{H}^+}$ can be derived from a simultaneous solution of
Eqs.~(\ref{bal1}) to (\ref{bal3}) as a
function of $P_{\mathrm{soil}}$. The sulfur content of the soil can be taken into account analogously.
The equilibrium constants for the chemical activities of the carbon and sulfur systems involved are
taken from \cite{stum81}. Note that the sulfur content of the soil also contributes to the
global weathering rate, but its influence does not depend on the temperature. It can be regarded as
an overall weathering attribute that has to be taken into account for the estimation of the
present-day value.

For any given weathering rate, the surface temperature and the CO$_2$ concentration of the
soil can be calculated in a self-consistent manner. $P_{\mathrm{soil}}$ is assumed to be
linearly related to the terrestrial biological productivity $\Pi$ \citep[see][]{volk87}
and the atmospheric CO$_2$ concentration $P_{\mathrm{CO}_2}$. Thus we find
\begin{equation}
\frac{P_{\mathrm{soil}}}{P_{\mathrm{soil},0}} \ = \ \frac{\Pi}{\Pi_0} \left( 1- \frac{P_{\mathrm{CO}_2,0}}{P_{\mathrm{soil},0}} \right)
     + \frac{P_{\mathrm{CO}_2}}{P_{\mathrm{soil},0}} ,\label{hz:eq2}
\end{equation}
where $P_{\mathrm{soil},0}$, $\Pi_0$ and $P_{\mathrm{CO}_2,0}$ are again present-day values.
Note that the present-day concentration of CO$_2$ in the soil is ten times the present-day
concentration of CO$_2$ in the atmosphere, i.e., $P_{\mathrm{soil}} = 10~P_{\mathrm{soil},0}$.

\subsection{Thermal Evolution Model}

Parameterized convection models are the simplest models for investigating the thermal
evolution of terrestrial planets and satellites. They have successfully been applied to the 
evolution of Mercury, Venus, Earth, Mars, and the Moon \citep{stev83,slee00}.
\cite{fran95} have investigated the thermal and volatile history of Earth and Venus in the
framework of comparative planetology. The internal structure of massive terrestrial planets
with one to ten Earth masses has been investigated by \cite{vale06} to
obtain scaling laws for the total radius, mantle thickness, core size, and average density as 
a function of mass.  Further scaling laws were found for different compositions. We will
use such scaling laws for mass-dependent properties of our 10~$M_{\oplus}$ super-Earth model
as well as for mass-independent material properties given by \cite{fran95}
(see Tab.~\ref{param}).

The thermal history and future of a super-Earth has to be determined to
calculate the spreading rate for solving Eq.~(\ref{gfr}).
A parameterized model of whole mantle convection including the volatile exchange
between the mantle and surface reservoirs \citep{fran95,fran98} is applied.
Assuming conservation of energy, the average mantle temperature $T_m$ can be
obtained by solving
\begin{equation} 
{4 \over 3} \pi \rho c (R_m^3-R_c^3) \frac{dT_m}{dt} \ = \ -4 \pi
R_m^2 q_m + {4 \over 3} \pi E(t) (R_m^3-R_c^3), \label{therm} \end{equation}
where $\rho$ is the density, $c$ is the specific heat at constant pressure,
$q_m$ is the heat flow from the mantle, $E(t)$ is the energy production rate by
decay of radiogenic heat sources in the mantle per unit volume, and $R_m$ and
$R_c$ are the outer and inner radii of the mantle, respectively. The radiogenic
heat source per unit volume is parameterized as
\begin{equation}
E(t) \ = \ E_0e^{-\lambda t}
\end{equation}
where $\lambda$ is the decay constant and the constant $E_0$
is obtained from the present heat flux of $q_m=0.07$ Wm$^{-2}$
for an Earth-size planet at 4.6 Gyr.

The mantle heat flow is parameterized in terms of the Rayleigh number $\mathrm{Ra}$ as
\begin{equation}
q_m \ = \ {k (T_m - T_{\mathrm{surf}}) \over R_m -R_c} \left({\mathrm{Ra} \over
\mathrm{Ra}_{\rm{crit}}}\right)^\beta \label{eqheat}
\end{equation}
with
\begin{equation}
\mathrm{Ra} \ = \ {g \alpha (T_m - T_{\mathrm{surf}}) (R_m - R_c)^3 \over \kappa \nu},
\label{eqrayleigh}
\end{equation}
where $k$ is the thermal conductivity, $\mathrm{Ra}_{\rm{crit}}$ is the critical value
of $\mathrm{Ra}$ for the onset of convection, $\beta$ is an empirical constant, $g$ is
the gravitational acceleration, $\alpha$ is the coefficient of thermal expansion,
$\kappa$ is the thermal diffusivity, and $\nu$ is the water-dependent
kinematic viscosity. The viscosity $\nu$ can be calculated with the help of a water
fugacity-dependent mantle creep rate. It strongly depends on the evolution of 
the mass of mantle water $M_w$, and the mantle temperature $T_m$, i.e.,
$\nu\equiv\nu(T_m,M_w)$ and is parameterized according to \cite{fran95}. 

The evolution of the mantle water can be described by a balance equation between
the regassing flux $F_{\mathrm{reg}}$ and outgassing flux $F_{\mathrm{out}}$ as
\begin{eqnarray}
\frac{dM_w}{dt} & \ = \ & F_{\mathrm{reg}}-F_{\mathrm{out}}  \nonumber \\
 & \ = \ & f_{\mathrm{bas}}\rho_{\mathrm{bas}}d_{\mathrm{bas}}SR_{\mathrm{H_2O}}-\frac{M_w}
         {\frac{4}{3}\pi(R_m^3-R_c^3)}d_mf_wS,
\label{eq:water}
\end{eqnarray} 
where
$f_{\mathrm{bas}}$ is the water content in the basalt layer,  
$\rho_{\mathrm{bas}}$ is the average density,
$d_{\mathrm{bas}}$ is the average thickness of the
basalt layer before subduction,
$S$ is the areal spreading rate,
$d_m$ is the melt generation depth and $f_w$ is the outgassing fraction of water.
$R_{\mathrm{H_2O}}$ is the regassing ratio of water, i.e., the fraction of subducting
water that actually enters the deep mantle.  The average thickness of the basalt layer
as well as the melt generation depth scale inversely with surface gravity $g$, i.e.,
$d_\mathrm{bas}\propto 1/g$ and $d_m\propto 1/g$.  The pressure closing of cracks in
the deeper parts of the basalt layer scales also inversely with $g$ and thus reduces
the storage capacity of volatiles for a super-Earth planet.  Therefore, the ratio
$F_\mathrm{reg}/F_\mathrm{out}$ is independent of $g$.  According to Eq.~(\ref{eq:water})
gravity influences only the time scale of mantle water evolution.  Therefore, as a
first approximation the melt generation depth $d_m$ does not depend on mantle
temperature.  However, there is a temperature dependence of $d_m$
\citep{mcke88,lang92}.  The regassing ratio depends linearly on the mean
mantle temperature $T_m$ that is derived from the thermal evolution model via
\begin{equation} 
R_{\mathrm{H_2O}}(T_m) \ = \ R_T \cdot\left(T_m(0)-T_m\right)+R_{\mathrm{H_2O},0}.\label{eq5} 
\end{equation}
The factor $R_T$ is adjusted to obtain the correct modern amount of surface water
(one ocean mass) for an Earth-size planet and $R_{\mathrm{H_2O},0}$ is fixed at $0.001$.
This value is obviously very low at the beginning of the planetary evolution because
of the enhanced loss of volatiles resulting from back-arc volcanism at higher temperatures.

The areal spreading rate $S$ is a function of the average mantle temperature $T_m$, the 
surface temperature $T_{\mathrm{surf}}$, the heat flow from the mantle $q_m$, and the
area of ocean basins $A_0$ \citep{turc82}, given as
\begin{equation}  S \ = \ \frac{q_m^2 \pi
\kappa A_0}{4 k^2 (T_m - T_\mathrm{surf})^2}\,. 
\end{equation}
In order to calculate the spreading rates for a planet with several Earth masses, 
the planetary parameters have to be adjusted accordingly.  We assume
\begin{equation}
\frac{R_p}{R_{\oplus}} \ = \ \left(\frac{M}{M_{\oplus}}\right)^{0.27}
\end{equation} and
with $R_p$ as planetary radius, see \citep{vale06}.
The total radius, mantle thickness, core size and average density are all functions
of mass, with subscript $\oplus$ denoting Earth values.
The exponent of $0.27$ has been obtained for super-Earths ($M>1 M_\oplus$), and has
already been used by \cite{bloh07b} in their models of Gliese 581c
and 581d.  The values of $R_m$, $R_c$, $A_0$, the density of the planet, and
the other planetary properties are also scaled accordingly.

The source of CO$_2$ to the atmosphere is expressed in mass of carbon outgassed at the
spreading zones, $C_{\mathrm{sr}}\propto S$. It has to be converted to an equivalent
concentration of CO$_2$ in the atmosphere. This can be done by the following equation
\begin{equation}
P_{\mathrm{CO}_2} \ = \
\frac{g}{4\pi R_p^2}\frac{\mu_{\mathrm{CO}_2}}{\mu_{\mathrm{C}}}C_{\mathrm{sr}},
\end{equation}
where $\mu_{\mathrm{CO}_2}$ and $\mu_{\mathrm{C}}$ are the molar weights of CO$_2$ and C,
respectively. The mass dependent pre-factor $g/R_p^2$ scales as $M^{-0.08}\approx M^0$
and has therefore been neglected in our study.  Therefore the conversion does not depend
on the planetary mass and the spreading rates $S$ can be directly used to calculate
$f_{\mathrm{sr}}$ in Eq.~(\ref{gfr}).

In Tab.~\ref{param} we give a summary of the selected values for the parameters used
in the thermal evolution model of the 10~$M_\oplus$ super-Earth planet, while also
depicting an Earth-size planet for comparison.  According to
\cite{vale07b}, we assume that a more massive planet is likely to convect in a plate
tectonic regime similar to Earth. Thus, the more massive the planet is, the higher the
Rayleigh number that controls convection, the thinner the top boundary layer (lithosphere),
and the faster the convective velocities.  This is the so-called boundary-layer limit
of convection.  From this limit it follows that the interior of a super-Earth is
always hotter and less viscous than that of a Earth-mass planet.  Nevertheless,
friction is the rate-limiting process for subduction.  Increasing the planetary
radius acts to decrease the ratio between driving forces and resistive strength
\citep{onei07}.  Thus a super-sized Earth might be in an episodic or stagnant
lid regime.

In a first order approximation, we assume a fixed thickness of the basalt layer and
melting depth corresponding to relatively low values.  Furthermore, the initial amount
of water $M_w(0)$ scales linearly with the planetary mass.  However, his might be an
underestimate because more massive planets tend to accrete more volatiles.


\clearpage

\section{Results}

\subsection{Habitability based on the integrated system approach}


\bigskip
\noindent
In the following, we study the habitability of super-Earth planets based on the
integrated system approach that has previously been used in various other planetary
studies \citep[e.g.,][]{fran00b,cunt03,bloh03,fran03,bloh07a,bloh07b}.  The
simulations have been carried out for a 10~$M_\oplus$ mass super-Earth with
a fixed relative continental area $c$ varied from $0.1$ to $0.9$.
Fig.~3 shows the behavior of the photosynthesis-sustaining habitable zone
(pHZ) of the Sun for a 10~$M_\oplus$ super-Earth planet. The age domain beyond
11~Gyr that also includes the post--main-sequence evolution is depicted in
Fig.~4.  The width of the pHZ during the main-sequence evolution is found to be
approximately constant, but for higher ages, it increases over time and moves
outward, a phenomenon most noticeable beyond 11.5~Gyr.  For example, for ages
of 11.0, 11.5, 12.0, and 12.1 Gyr, the pHZ is found to extend from 1.41 to 2.60,
1.58 to 2.60, 4.03 to 6.03, and 6.35 to 9.35 AU, respectively.

At relatively high ages, habitable solutions are identified as water worlds,
if the Sun as central star has reached the RGB. The reason is that planets
with a considerable continental area have higher weathering rates that provide
the main sink of atmospheric CO$_2$.  Therefore, such planets are unable to build
up CO$_2$-rich atmospheres which prevent the planet from freezing or allowing
photosynthesis-based life.  This result is consistent with previous findings
for Earth-mass planets around the Sun or stars of similar masses \citep{fran00b,
cunt03}.

Note that the partial pressure of carbon dioxide in the planetary atmosphere
is determined by the equilibrium of sources and sinks.  The sources are given
by volcanic outgassing, while the sinks are given by the weathering of silicates
on the continents.  As previously found in studies of 1~$M_\oplus$ planets
\citep[e.g.,][]{fran00b}, the rate of outgassing is monotonously decreasing
with age because of the decay of long-lived radionuclides and the loss of the
initially available accretion energy at the planetary surface.  This process
starts just after completion of the planetary accretion both for an initially
habitable and uninhabitable planet.  A planet beyond the outer edge of the pHZ
is completely frozen and thus no weathering will occur on the continents.
Therefore, all CO$_2$ is accumulated in the atmosphere.  If the planet becomes
habitable due to the increase of the luminosity of the central star, weathering
starts and a new equilibrium of atmospheric CO$_2$ is established as a consequence.

Furthermore, the interior of a planet with a relatively low mass is known to
cool down more rapidly.  Therefore, such a planet initially beyond the outer edge
of the habitable zone will not become habitable at a later stage because of
the failure to provide a sufficiently dense atmosphere.  In contrast, a
super-Earth planet might become habitable, depending on the relative size
of the continental area.  In a recent study, the importance of snowball planets
as a possible source of water-rich terrestrial planets was elucidated by
\cite{taji08}, although the main focus of this paper was the assessment of
internal oceans.

Super-Earth-type water worlds are even able to realize pHZ-type
habitability beyond solar-type main-sequence evolution.  Any model where
mantle vents its water will end up as a water world super-Earth. The height
of ridges, volcanos, and mountains scale with lithosphere thickness and
hence with $1/g$.  As the central star
evolves, its pHZ expands outward, and moves further away from
the star, particularly for stellar ages beyond 11.8~Gyr (see Fig.~4).
Similar to the climatic HZ (see \citeauthor{lope05} for details), the pHZ
acts like a shell that sweeps progressively outward over a wide range of
distances from the star.  This results in a significant decrease of the
duration of the transit of the habitable zone for any planet located
beyond 1.5~AU (see Fig.~5).  We find that for water worlds with $c=0.1$,
the duration of the transit of the pHZ at 2, 3, and 5~AU is
given as 3.7, 0.25, and 0.10~Gyr, respectively, whereas for planets at
10 and 20~AU, much smaller durations of the transit are identified.

Figs.~3 and 4 also depict various orbital distances of planets
originally located between 1~AU and 5~AU.  Note that these orbital distances do
not change during the stellar main-sequence stage, i.e., below 10~Gyr \citep{schr08},
owing to the lack of significant mass loss and in the absence of significant
planet-planet interaction as typically encountered in multiple planetary systems.
Thereafter, the orbital distances $R$ of any planet increases following
$R \propto M_\odot^{-1}$ with $M_\odot$ as mass owing to the conservation of
planetary orbital angular momentum.

\subsection{Comparison with previous results}

The existence of habitability around stars that have evolved away from the
main-sequence has already been the topic of previous investigations.
\cite{lope05} studied the behavior of the climatic HZ based on the
concept of \cite{kast93} for stars of different masses, including the
Sun.  \cite{lope05} assume a HZ based on conservative
limits of habitability.  The inner limit of their conservative estimate
is set by the lowest temperature at which the liquid-solid phase change
of water occurs.  The estimate of the outer limit assumes the existence
of a greenhouse effect involving CO$_2$ and H$_2$O gas \citep{kast93}.
The less conservative definition extends the outer edge of the limit of
habitability to as large as 2.4~AU, largely depending on the radiative
properties of the CO$_2$ ice clouds; see \cite{forg97} and \cite{misc00}
for detailed studies.

Akin to the pHZ previously discussed, \cite{lope05} found that for the
Sun during its evolution the climatic HZ acts like a shell that sweeps
progressively outward over a wide range of distances from the star.  The
duration of the transit during which the HZ passes over a planet located
at 1~AU from the star was found to be on the order of 10$^9$~yr.  After
the star leaves the main-sequence, the climatic HZ progressively moves to
2~AU.  The duration of the transit at this location is approximately
10$^9$~yr.  A plateau is observed in the curve up to 9~AU (for the
conservative limits) and up to 13~AU (for the less conservative limits),
where the durations of habitable conditions lasts from a few to several
times 10$^8$~yr.  At 10~AU, the duration is smaller, about 10$^8$~yr.
At 15~AU from the star the duration of habitable conditions lasts more
than 10$^7$ yr, and at the largest distances considered in the study by
\cite{lope05} the duration gradually decreases.

Note that the model of solar evolution considered in the \citeauthor{lope05}
study is that by \cite{maed88}.  Nonetheless, their results would be
quite similar if they had used the subsequent model by \cite{sack93}
or the very recent model by \cite{schr08}.  The reason is that the outcome
of the \citeauthor{lope05} study is much more dependent on the choices
made concerning the upper and lower limits of the climatic HZ, mostly
connected to the treatment of the CO$_2$ atmospheres (i.e., radiative
properties, cloud coverage, etc.) than to the adopted model of solar
evolution.

Fig.~6 shows the comparison between the work by \cite{lope05}
and our current results.   We find that for water worlds ($c=0.1$),
the transit times for photosynthesis-sustaining habitability (pHZ) for
planets at a given reference distance from the star is relatively
similar to the results obtained for the conservative climatic HZ
\citep{kast93} adopted by \citeauthor{lope05}, albeit the transit
times in our study are typically lower by a factor of up to 1.5.
For example, the durations of the transit  concerning pHZ-type
habitability for water worlds at 2, 3, and 5~AU are identified
as 3.7, 0.25, and 0.10~Gyr, respectively, whereas for planets at
10 and 20~AU, the durations of the transit found are as low as
27 and 9~Myr, respectively.

However, especially at distances beyond 2~AU, significantly smaller transit
times are encountered for planets with larger continental areas
in terms of all stellar distances, a result consistent with previous
findings.  For $c=0.5$, the transit time of the pHZ drops beneath
1~Gyr for planets located at 1.8~AU.  For planets with a relative
continental area of $c=0.9$, also referred to as ``land worlds'',
no significant photosynthesis-sustaining habitability is found for
planets beyond 1.5~AU.

\section{Summary and Conclusions}

We studied the habitability of super-Earth planets based on the integrated
system approach that has previously been used in various theoretical planetary
studies \cite[e.g.,][]{fran00b,cunt03,bloh03,fran03,bloh07a,bloh07b}.  This
work is motivated by the quest to identify habitability outside the
Solar System as well as the ongoing discovery of super-Earths in the solar
neighborhood with the Gliese 876 \citep{rive05} and Gliese 581 \citep{udry07}
systems as prime examples. 

In agreement with previous studies, it is found that photosynthesis-sustaining
habitability strongly depends on the planetary characteristics.  For planets
of a distinct size, the most important factor is the relative continental area. 
Habitability was found
most likely for water worlds, i.e., planets with a relatively small
continental area.  For planets at a distinct distance from the central star,
we identified maximum durations of the transit of the pHZ.  A comparison of planets
with different masses revealed that the maximum duration of the transit increases with
planetary mass.  Therefore, the upper limit for the duration of the transit for
any kind of Earth-type planet is found for most massive super-Earth planets, i.e.,
10~$M_\oplus$, rather than 1~$M_\oplus$ planets, which are rendered
uninhabitable after 6.5~Gyr, as previously pointed out by \cite{fran00b}.

Our study forwards a thermal evolution model for a 10~$M_\oplus$ super-Earth
orbiting a star akin to the Sun.  The calculations consider updated models
of solar evolution obtained by \cite{schr08} with a detailed mass loss
description provided by \cite{schr05}.  The latter is relevant for the
change of luminosity along the Red Giant Branch as well as the increase
of the orbital distances of any putative planets during that phase.  By
employing the integrated system approach, we were able to identify the sources
and sinks of atmospheric carbon dioxide on the planet, allowing us to describe
the photosynthesis-sustaining habitable zone (pHZ) determined by the limits
of biological productivity on the planetary surface.

Concerning the pHZ, we identified the following properties:

\smallskip
\noindent
(1) Geodynamic solutions are identified for different solar ages, including the
RGB phase.  The pHZ increases in width over time and moves outward.  For example,
for ages of 11.0, 11.5, 12.0, and 12.1 Gyr, the pHZ is found to extend from
1.41 to 2.60, 1.58 to 2.60, 4.03 to 6.03, and 6.35 to 9.35 AU, respectively.

\smallskip
\noindent
(2) Habitable solutions at large ages, especially for the subgiant and giant
phase, are water worlds.  This also means that the possibility of water worlds in
principle results in an extension of the outer edge of habitability.  The reason
is that planets with a considerable continental area have higher weathering rates
that provide the main sink of atmospheric CO$_2$.  Therefore, such planets,
contrary to water worlds, are unable to build up CO$_2$-rich atmospheres that
prevent the planet from freezing or allowing photosynthesis-based life.

\smallskip
\noindent
(3) The total duration of the transit of the habitable zone is similar to the
predictions by \cite{lope05} based on the conservative limits of the climatic
HZ obtained by \cite{kast93}.  For water worlds with $c=0.1$, the transit times
of the pHZ at 2, 3, and 5~AU are obtained as 3.7, 0.25, and 0.10~Gyr, respectively,
whereas for planets at 10 and 20~AU, much smaller transit times are found.

\medskip

Our results are a further motivation to consider super-Earth planets in
upcoming or proposed planet search missions such as Kepler, TPF or Darwin.
Moreover, our results can also be viewed as a reminder to seriously
contemplate the possibility of habitable planets around red giants, as
previously pointed out by \cite{lope05} and others.  For central stars with
a higher mass than the Sun, a more rapid evolution will occur that will also
affect the temporal and spatial constraints on planetary habitability
when the central stars have reached the RGB.

\noindent {\bf Acknowledgments}

We would like to thank Norman Sleep and an anonymous referee for their helpful
comments which allowed us improving the paper. 

\pagebreak

\newpage

\afterpage{\clearpage
\begin{figure}
\centering
\includegraphics[height=10cm]{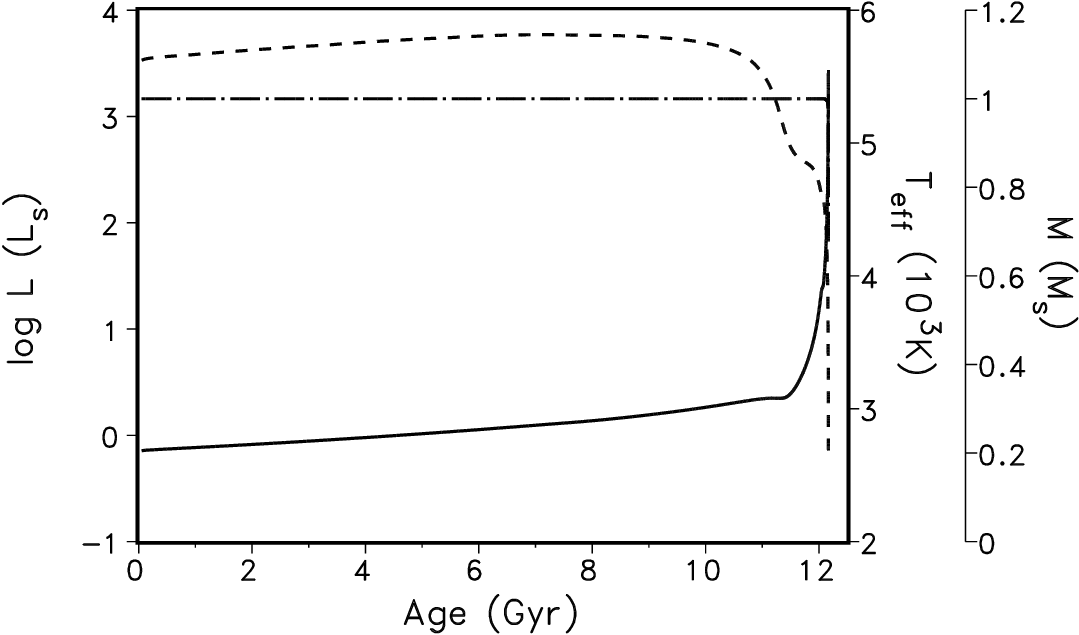}
\caption{Solar evolution model following \cite{schr08},
depicting the luminosity (solid line), the effective temperature
(dashed line), and the mass (dash-dotted line).
}
\bigskip
\label{lumin}
\end{figure}}

\afterpage{\clearpage
\begin{figure}
\centering
\includegraphics[height=10cm]{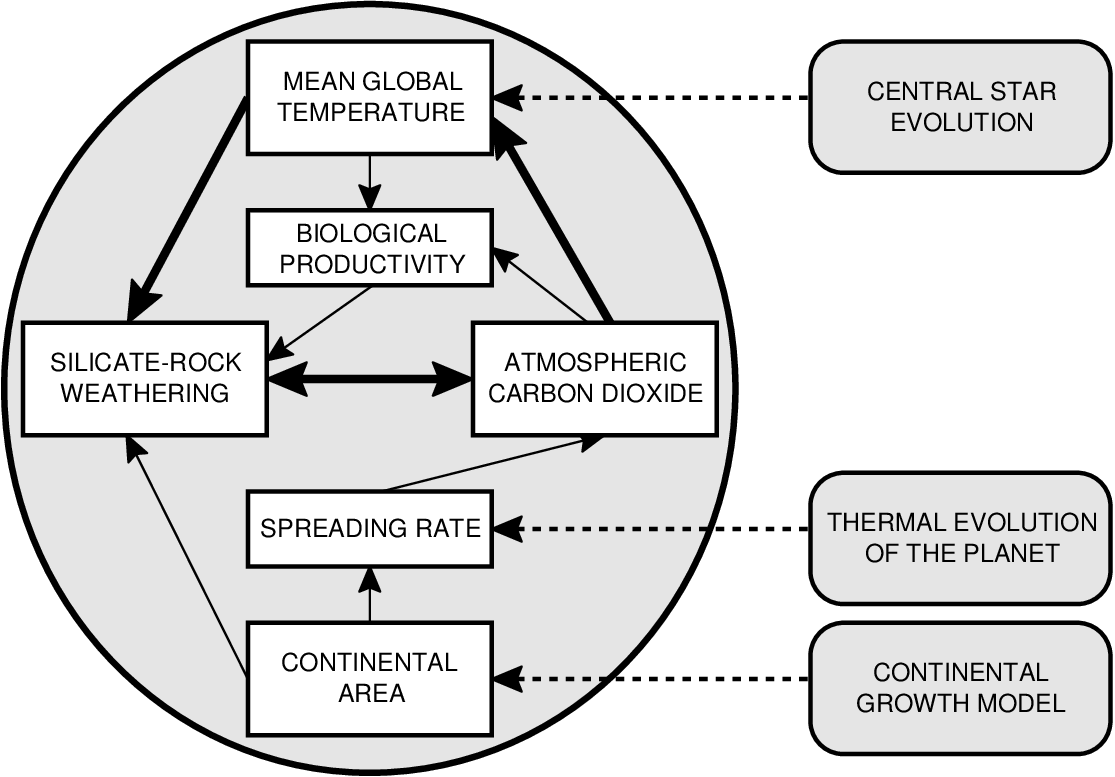}
\caption{Box model of the integrated system approach adopted in
our model.  The arrows indicate the different types of forcing,
which are the main feedback loop for stabilizing the climate
(thick solid arrows), the feedback loop within the system
(thin solid arrows), and the external and internal forcings
(dashed arrows).
}
\bigskip
\label{boxmodel}
\end{figure}}

\afterpage{\clearpage
\begin{figure}
\centering
\includegraphics[height=10cm]{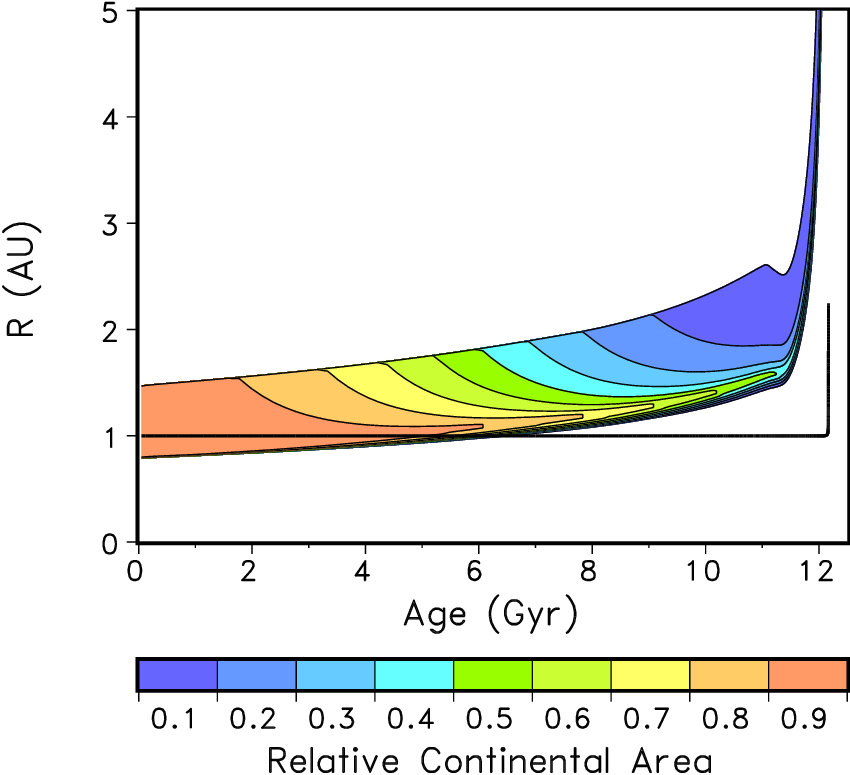}
\caption{The photosynthesis-sustaining habitable zone (pHZ) of the Sun
for a super-Earth planet also considering the solar evolution along the
Red Giant Branch.  The colored areas indicate the extent of the pHZ for
different relative continental areas.  The solid line depicts the orbital
distance of a planet originally located at 1.0 AU. 
}
\bigskip
\label{hz1}
\end{figure}}

\afterpage{\clearpage
\begin{figure}
\centering
\includegraphics[height=10cm]{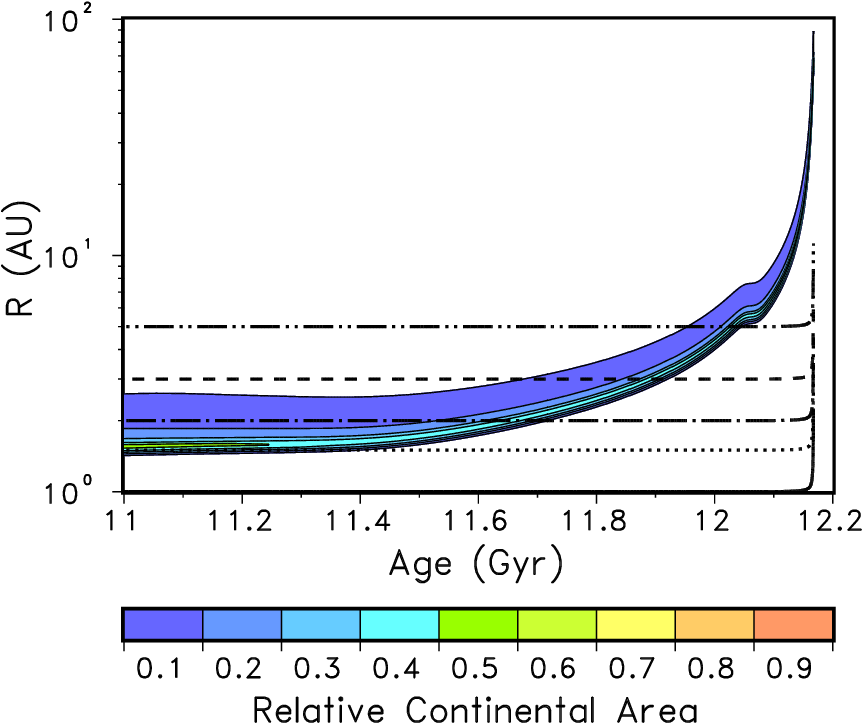}
\caption{Same as Fig.~\ref{hz1}, but now zoomed-in at the
Red Giant Branch evolution beyond 11 Gyr.  We also show the
distance evolution of various fictitious planets, originally
located at 1.0 AU (solid line), 1.5 AU (dotted line),
2.0 AU (dash-dotted line), 3.0 AU (dashed line), and
5.0 AU (dash-double dotted line), respectively.
}
\bigskip
\label{hz2}
\end{figure}}

\afterpage{\clearpage
\begin{figure}
\centering
\includegraphics[height=15cm]{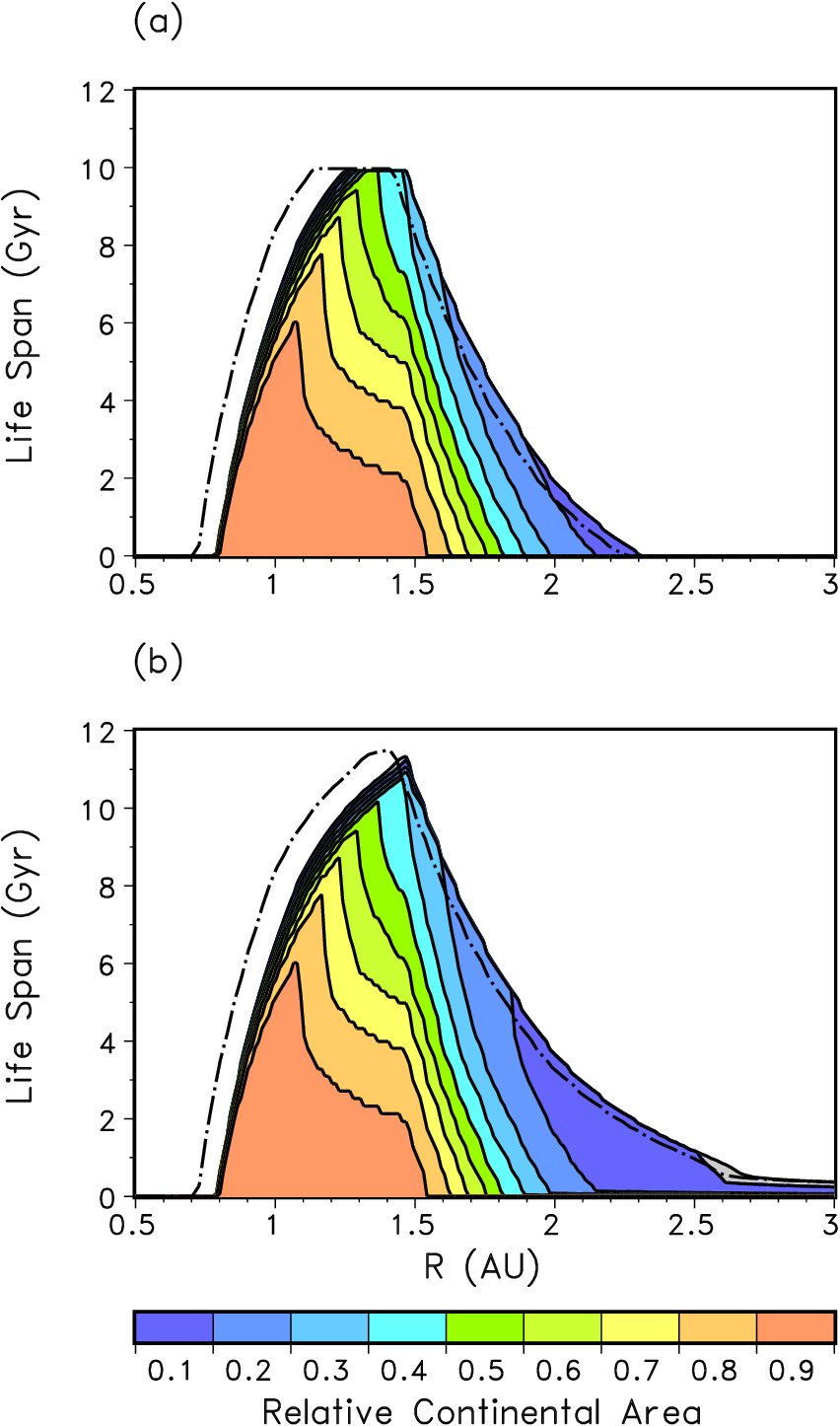}
\caption{Duration of the transit of the pHZ for a super-Earth as function
of the distance from the central star $R$ for different relative
continental areas.  Model (a) only includes the stellar
main-sequence evolution (10 Gyr), whereas Model (b) also includes
the evolution along the Red Giant Branch.  The gray area indicates
the result based on the geostatic approximation \citep[e.g.,][]{fran00b}
that is mostly enveloping the colored area.  The dash-dotted line
indicates transit times of the general habitable zone given by the model of
\cite{kast93}.
}
\bigskip
\label{lifespan}
\end{figure}}

\afterpage{\clearpage
\begin{figure}
\centering
\includegraphics[height=10cm]{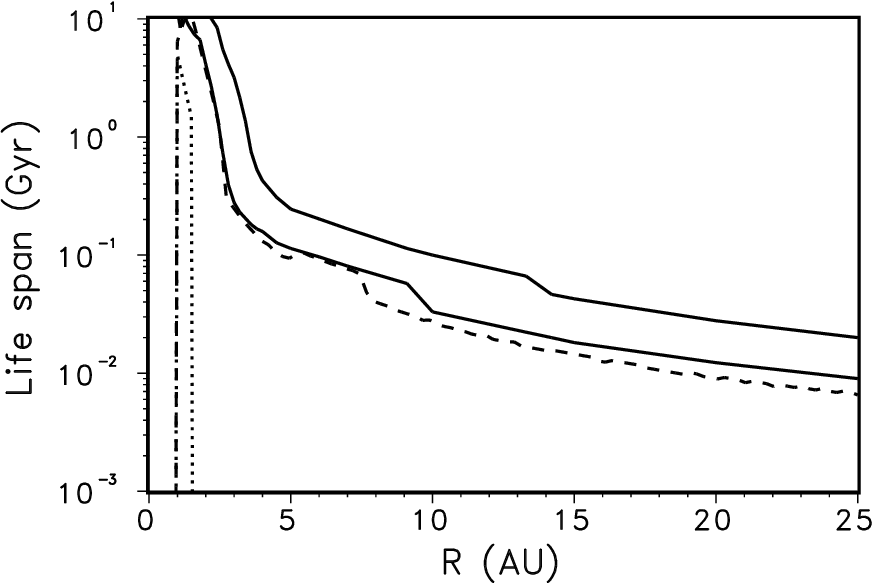}
\caption{Duration of the transit of the pHZ for super-Earth planets as function
of the planetary distance $R$ from the center of the star.  We depict the
results for ``water worlds'' ($c=0.1$; dashed line) and ``land worlds''
($c=0.9$; dotted line).  In addition, we also show the results by \cite{lope05}
for terrestrial planets (solid lines) based on the concept of habitability by
\cite{kast93}.  Here the lower line refers to the duration of habitability based
on their conservative definition of the HZ, whereas for the upper line the
less conservative definition has been adopted for the outer limit of the HZ,
while the inner limit of the HZ was left unchanged. 
}
\bigskip
\label{lopez}
\end{figure}}

\newpage
\renewcommand{\baselinestretch}{1.7}

\begin{table*}
\caption{Parameter values for the evolution model of the mantle temperature and water}             
\label{param}      
\centering                          

\begin{tabular}{l l l l l}        
\hline\hline                      
Parameter & \multicolumn{2}{c}{Value}   & Unit & Description   \\    
 ...      & $1~M_\oplus$ & $10~M_\oplus$  & ...  & ...         \\
\hline                            
$d_{\mathrm{bas}}$    & $5\times 10^3$ & $5\times 10^3$ & m & Average thickness of the basalt layer \\
$f_{\mathrm{bas}}$    & $0.03$ & $0.03$ & ... & Mass fraction of water in the basalt layer \\
$\rho_{\mathrm{bas}}$ & $2,950$ &  4,569 & kg m$^{-3}$ & Density of the basalt \\
$f_w$                 & $0.194$ & $0.194$ & ... & Degassing fraction of water \\
$d_m$                 & $40\times 10^3$  & $40\times 10^3$ & m & melting depth \\
$k$                   & $4.2$ & $4.2$ &  J s$^{-1}$ m$^{-1}$ K$^{-1}$ & Thermal conductivity \\
$R_c$                 & $3,471\times 10^3$ &  $6,463\times 10^3$ & m & Inner radius of the mantle \\
$R_m$                 & $6,271\times 10^3$ & $11,667\times 10^3$ & m & Outer radius of the mantle \\
$M_w(0)$              & $4.2\times 10^{21}$ &  $14.04\times 10^{22}$ & kg & Initial amount of mantle water \\
$T_m(0)$              & $3,000$ & $3,000$ & K & Initial mantle temperature \\
$K_{\rm H}$           & $3.36\times 10^{-4}$    &  $3.36\times 10^{-4}$   & mol J$^{-1}$ & Henry's law constant at 25$^\circ$C \\
$\log K_1$                 & -6.3    &  -6.3   & ... & Equilibrium constant in Eq.~(\ref{bal1}) at 25$^\circ$C\\
$\log K_2$                 & -10.3    &  -10.3   & ... & Equilibrium constant in Eq.~(\ref{bal2}) at 25$^\circ$C \\
$\kappa$              & $10^{-6}$ & $10^{-6}$ & m$^2$ s$^{-1}$ & Thermal diffusivity \\
$\rho c$              & $4.2\times 10^6$ &  $4.2\times 10^6$ & J m$^{-3}$ K$^{-1}$ & Density $\times$ specific heat \\
$R_T$                 & $2.98\times 10^{-4}$ &  $2.98 \times 10^{-4}$ & K$^{-1}$ & Temperature dependence of regassing ratio \\
$\alpha$              & $3\times 10^{-5}$ & $3\times 10^{-5}$ & K$^{-1}$ & Coefficient of thermal expansion \\
$\beta$               & $0.3$ & $0.3$  & ... & Empirical constant in Eq.~(\ref{eqheat}) \\
$\mathrm{Ra}_{\mathrm{crit}}$ & $1,100$ & $1,100$ &  ... & Critical Rayleigh number \\
$\lambda$             & $0.34$ & $0.34$&  Gyr$^{-1}$ & Decay constant \\
$E_0$                 & $1.46\times 10^{-7}$ &  $1.46\times 10^{-7}$ & J s$^{-1}$ m$^{-3}$ & Initial heat generation per time and volume \\
$g$                   & 9.81 & 28.26 & m s$^{-2}$ & Gravitational acceleration \\
\hline                                   
\end{tabular}
\end{table*}


\begin{thebibliography}{}

\bibitem[Buccino {\em et al.}(2006)]{bucc06}
Buccino, A.P., Lemarchand, G.A., and Mauas, P.J.D. (2006) Ultraviolet radiation
constraints around the circumstellar habitable zones. {\em Icarus} 183, 491--503.

\bibitem[Chyba and Hand(2005)]{chyb05}
Chyba, C.F. and Hand, K.P. (2005) Astrobiology: The study of the living universe.
{\em Ann. Rev. Astron. Astrophysics} 43, 31--74.

\bibitem[Cuntz {\em et al.}(2003)]{cunt03}
Cuntz, M., von Bloh, W., Bounama, C., and Franck, S. (2003) On the possibility of
Earth-type habitable planets around 47 UMa. {\em Icarus} 162, 214--221.

\bibitem[Cuntz {\em et al.}(2008)]{cunt08}
Cuntz, M., Gurdemir, L., Guinan, E.F., and Kurucz, R.L. (2008) Astrobiological
effects of F, G, K and M main-sequence stars. In Exoplanets:
Detection, Formation \& Dynamics, Proc. IAU Symposium 249, ed. Y.-S. Sun,
S. Ferraz-Mello, and J.-L. Zhou (Cambridge: Cambridge University Press), 203--206.

\bibitem[Forget and Pierrehumbert(1997)]{forg97}
Forget, F. and Pierrehumbert, R.T. (1997) Warming early Mars with carbon dioxide
clouds that scatter infrared radiation. {\em Science} 278, 1273--1276.

\bibitem[Franck(1998)]{fran98}
Franck, S. (1998) Evolution of the global mean heat flow over 4.6 Gyr.
{\em Tectonophysics} 291, 9-18.

\bibitem[Franck {\em et al.}(2000a)]{fran00a}
Franck, S., Block, A., von Bloh, W., Bounama, C., Schellnhuber, H.-J., and
Svirezhev, Y. (2000a) Reduction of life span as a consequence of geodynamics.
{\em Tellus} 52B, 94--107.

\bibitem[Franck {\em et al.}(2000b)]{fran00b}
Franck, S., von Bloh, W., Bounama, C., Steffen, M., Sch\"onberner, D., and
Schellnhuber, H.-J. (2000b) Determination of habitable zones in extrasolar
planetary systems: Where area Gaia's sisters? \jgr 105 (E1), 1651--1658.

\bibitem[Franck {\em et al.}(2003)]{fran03}
Franck, S., Cuntz, M., von Bloh, W., and Bounama, C. (2003) The habitable zone
of Earth-mass planets around 47 UMa: Results for land and water worlds. \ija 2,
35--39.

\bibitem[Franck and Bounama(1995)]{fran95}
Franck, S. and Bounama, C. (1995) Effects of water-dependent creep rate
on the volatile exchange between mantle and surface reservoirs.
{\em Phys. Earth Planet. Inter.} 92, 57--65.

\bibitem[Gough(1981)]{goug81}
Gough, D.O. (1981) Solar interior structure and luminosity variations.
\mnras 74, 21--34.

\bibitem[Guinan and Ribas(2002)]{guin02}
Guinan, E.F. and Ribas, I. (2002) Our changing Sun: The role of solar nuclear
evolution and magnetic activity on Earth's atmosphere and climate.
In ASP Conf. Ser. 269, The Evolving Sun and Its Influence on Planetary Environments,
ed. B. Montesinos, A. Gimenez and E.F. Guinan (San Francisco: ASP), 85--106.

\bibitem[Jones {\em et al.}(2005)]{jone05}
Jones, B.W., Underwood, D.R., and Sleep, P.N. (2005) Prospects for habitable
``Earths'' in known exoplanetary systems. \apj 622, 1091--1101.

\bibitem[Kasting(1988)]{kast88}
Kasting, J.F. (1988) Runaway and moist greenhouse atmospheres and the evolution of
Earth and Venus. {\em Icarus} 74, 472--494.

\bibitem[Kasting and Ackerman(1986)]{kast86}
Kasting, J.F. and Ackerman, T.P. (1986) Climatic consequences of very high carbon
dioxide levels in the Earth's early atmosphere. {\em Science} 234, 1383--1385.

\bibitem[Kasting {\em et al.}(1993)]{kast93}
Kasting, J.F., Whitmire, D.P., and Reynolds, R.T. (1993) Habitable zones around
main sequence stars. {\em Icarus} 101, 108--128.

\bibitem[Langmuir {\em et al.}(1992)]{lang92}
Langmuir, C.H., Klein, E.M., and Plank, T. (1992) Petrological systematics of
mid-ocean ridge basalts: Constraints on melt generation beneath ocean ridges.
{\em AGU Monograph} 71, 183--280.

\bibitem[Lineweaver {\em et al.}(2004)]{line04}
Lineweaver, C.H., Fenner, Y., and Gibson, B.K. (2004) The galactic habitable zone
and the age distribution of complex life in the Milky Way. {\em Science} 303, 59--62.

\bibitem[Lopez {\em et al.}(2005)]{lope05}
Lopez, B., Schneider, J., and Danchi, W.C. (2005) Can life develop in the expanded
habitable zones around red giant stars? \apj 627, 974--985.

\bibitem[Maeder and Meynet(1988)]{maed88}
Maeder, A. and Meynet, G. (1988) Tables of evolutionary star models from 0.85 to
120 M$_\odot$ with overshooting and mass loss. \aaps 76, 411--425.  

\bibitem[McKenzie and Bickle(1988)]{mcke88}
McKenzie, D. and Bickle, M. J. (1988) The volume and composition of melt generated
by extension of the lithosphere. {\em J. Petrology} 29, 625--679.

\bibitem[Mischna {\em et al.}(2000)]{misc00}
Mischna, M.A., Kasting, J.F., Pavlov, A., and Freedman, R. (2000) Influence of
carbon dioxide clouds on early martian climate. {\em Icarus} 145, 546--554.

\bibitem[O'Neill and Lenardic(2007)]{onei07}
O'Neill, C. and Lenardic, A. (2007) Geological consequences of super-sized Earths.
\grl 34, L19204, doi:10.1029/2007GL030598.

\bibitem[Pols {\em et al.}(1995)]{pols95}
Pols, O.R., Tout, C.A., Eggleton, P.P., and Han, Z. (1995) Approximate input
physics for stellar modelling. \mnras 274, 964--974.

\bibitem[Pols {\em et al.}(1998)]{pols98}
Pols, O.R., Schr\"oder, K.-P., Hurley, J.R., Tout, C.A., and Eggleton, P.P.
(1998) Stellar evolution models for Z = 0.0001 to 0.03. \mnras 298, 525--536.

\bibitem[Rivera {\em et al.}(2005)]{rive05}
Rivera, E.J., Lissauer, J.J., Butler, R.P., Marcy, G.W., Vogt, S.S., Fischer, D.A.,
Brown, T.M., Laughlin, G., and Henry, G.W. (2005) A $\sim 7.5 M_\oplus$ planet
orbiting the nearby star, GJ 876. \apj 634, 625--640.

\bibitem[Sackmann {\em et al.}(1993)]{sack93}
Sackmann, I.-J., Boothroyd, A.I., and Kraemer, K.E. (1993) Our Sun.
III. Present and future. \apj 418, 457--468. 

\bibitem[Schwartzman(1999)]{schw99}
Schwartzman, D.W. (1999)  Life, Temperature, and The Earth: The Self-organizing
Biosphere. Columbia University Press, New York.

\bibitem[Schr\"oder and Cuntz(2005)]{schr05}
Schr\"oder, K.-P. and Cuntz, M. (2005)  A new version of Reimers' law of mass loss
based on a physical approach. \apjl 630, L73--L76.

\bibitem[Schr\"oder and Cuntz(2007)]{schr07}
Schr\"oder, K.-P. and Cuntz, M. (2007) A critical test of empirical mass loss formulas
applied to individual giants and supergiants. \aap 465, 593--601.

\bibitem[Schr\"oder and Smith(2008)]{schr08}
Schr\"oder, K.-P. and Smith, R.C. (2008) Distant future of the Sun and Earth revisited.
\mnras 386, 155--163.

\bibitem[Segura {\em et al.}(2003)]{segu03}
Segura, A., Krelove, K., Kasting, J.F., Sommerlatt, D., Meadows, V., Crisp, D.,
Cohen, M., and Mlawer, E. (2003) Ozone concentrations and ultraviolet fluxes
on Earth-like planets around other stars. {\em Astrobiology} 3, 689--708.

\bibitem[Selsis {\em et al.}(2007)]{sels07}
Selsis, F., Kasting, J.F., Levrard, B., Paillet, J, Ribas, I., and Delfosse, X.
(2007) Habitable planets around the star Gliese 581? \aap 476, 1373--1387.

\bibitem[Sleep(2000)]{slee00}
Sleep, N.H. (2000) Evolution of the mode of convection within terrestrial planets.
\jgr 105, 17563--17578.

\bibitem[Stevenson {\em et al.}(1983)]{stev83}
Stevenson, D.J., Spohn, T., and Schubert, G. (1983) Magnetism and thermal evolution
of the terrestrial planets. {\em Icarus} 54, 466--489.

\bibitem[Stumm and Morgan(1981)]{stum81}
Stumm, W. and Morgan, J.J. (1981) Aquatic Chemistry. An Introduction Emphasizing
Chemical Equilibria in Natural Waters. 2nd Ed., Wiley, New York.

\bibitem[Tajika(2008)]{taji08}
Tajika, E. (2008) Snowball planets as a possible type of water-rich terrestrial planet
in extrasolar planetary systems. \apjl 680, L53--L56.

\bibitem[Tajika and Matsui(1992)]{taji92}
Tajika, E. and  Matsui, T. (1992) Evolution of terrestrial proto-CO$_2$ atmosphere
coupled with thermal history of the Earth. {\em Earth Planet. Sci. Lett.}
113, 251--266.

\bibitem[Turcotte and Schubert(1982)]{turc82}
Turcotte, D.L. and Schubert, G. (1982) Geodynamics. Cambridge University Press,
Cambridge.

\bibitem[Udry {\em et al.}(2007)]{udry07}
Udry, S., Bonfils, X., Delfosse, X., Forveille, T., Mayor, M., Perrier, C.,
Bouchy, F., Lovis, C., Pepe, F., Queloz, D., and Bertaux, J.-L. (2007) The HARPS
search for southern extra-solar planets: XI. Super-Earths (5 and 8 $M_\oplus$)
in a 3-planet system. \aap 469, L43--L47.

\bibitem[Underwood {\em et al.}(2003)]{unde03}
Underwood, D.R., Jones, B.W., and Sleep, P.N. (2003) The evolution of habitable zones
during stellar lifetimes and its implications on the search for extraterrestrial life.
\ija 2, 289--299.

\bibitem[Valencia {\em et al.}(2006)]{vale06}
Valencia, D., O'Connell, R.J., and Sasselov, D. (2006) Internal structure of massive
terrestrial planets. {\em Icarus} 181, 545--554.

\bibitem[Valencia {\em et al.}(2007a)]{vale07a}
Valencia, D., Sasselov, D.D., and O'Connell, R.J. (2007a) Radius and structure models
of the first super-Earth planet. \apj 656, 545--551.

\bibitem[Valencia {\em et al.}(2007b)]{vale07b}
Valencia, D., O'Connell, R.J., and Sasselov, D.D. (2007b) Inevitability of plate
tectonics on super-Earths. \apjl 670, L45--L48.

\bibitem[Volk(1987)]{volk87}
Volk, T. (1987) Feedbacks between weathering and atmospheric CO$_2$ over the
last $100$ million years. {\em Am. J. Sci.} 287, 763--779.

\bibitem[von Bloh {\em et al.}(2003)]{bloh03}
von Bloh, W., Cuntz, M., Franck, S., and Bounama, C. (2003) On the possibility
of Earth-type habitable planets in the 55~Cancri system. {\em Astrobiology} 3,
681--688.

\bibitem[von Bloh {\em et al.}(2007a)]{bloh07a}
von Bloh, W., Bounama, C., and Franck, S. (2007a) Dynamic habitability for Earth-like
planets in 86 extrasolar planetary systems. {\em Planet. Space Sci.} 55, 651--660.

\bibitem[von Bloh {\em et al.}(2007b)]{bloh07b}
von Bloh, W., Bounama, C., Cuntz, M., and Franck, S. (2007b) The habitability of
super-Earths in Gliese 581. \aap 476, 1365--1371.

\bibitem[Walker {\em et al.}(1981)]{walk81}
Walker, J.C., Hays, P.B., and Kasting, J.F. (1981) A negative feedback mechanism
for the long-term stabilization of Earth's surface temperature. \jgr 86, 9776--9782.

\bibitem[Williams(1998)]{will98}
Williams, D.M. (1998) The stability of habitable planetary environments. Ph. D. thesis,
Pennsylvania State University.

\end{thebibliography}
\end{document}